\documentclass[twocolumn,showpacs,pre,superscriptaddress,floatfix]{revtex4}

\usepackage{color} 
\usepackage{tabularx} 
\usepackage{epsfig}
\usepackage{amsmath} 
\usepackage{amssymb} 
\usepackage{graphicx}
\usepackage{wasysym}
\usepackage{oldgerm}
\usepackage{psfrag}
\usepackage{times}

\begin{document}

\title{Charge avalanches and depinning in the Coulomb glass: The role of
long-range interactions}

\author{Juan Carlos Andresen}
\affiliation{Theoretische Physik, ETH Zurich, CH-8093 Zurich, Switzerland}
\affiliation{Department of Theoretical Physics, Royal Institute of Technology, Stockholm, Sweden}

\author{Yohanes Pramudya}
\affiliation{Department of Physics and National High Magnetic Field
Laboratory, Florida State University, Tallahassee, Florida 32306, USA}

\author{Helmut G.~Katzgraber}
\affiliation{Department of Physics and Astronomy, Texas A\&M University,
College Station, Texas 77843-4242, USA}
\affiliation{Santa Fe Institute, 1399 Hyde Park Road, Santa Fe, New
Mexico 87501 USA}
\affiliation{Applied Mathematics Research Centre, Coventry University,
Coventry, CV1 5FB, United Kingdom}

\author{Creighton K.~Thomas}
\affiliation {Department of Materials Science and Engineering, Northwestern
University, Evanston, Illinois 60208-3108, USA}

\author{Gergely T.~Zimanyi}
\affiliation{Department of Physics, University of California, Davis,
California 95616, USA}

\author{V.~Dobrosavljevi{\'{c}}}
\affiliation{Department of Physics and National High Magnetic Field
Laboratory, Florida State University, Tallahassee, Florida 32306, USA}

\date{\today}
\begin{abstract}

We explore the stability of far-from-equilibrium metastable states of a
three-dimensional Coulomb glass at zero temperature by studying charge
avalanches triggered by a slowly varying external electric field.
Surprisingly, we identify a sharply defined dynamical (``depinning'')
phase transition from stationary to nonstationary charge displacement at
a critical value of the external electric field.  Using
particle-conserving dynamics, scale-free system-spanning avalanches are
observed only at the critical field. We show that the qualitative
features of this depinning transition are completely different for an
equivalent short-range model, highlighting the key importance of
long-range interactions for nonequilibrium dynamics of Coulomb glasses.

\end{abstract}

\pacs{75.50.Lk, 75.40.Mg, 05.50.+q, 64.60.-i}

\maketitle

\section{Introduction}

The long-range nature of the Coulomb interaction plays only a secondary
role in metals, where it remains screened by mobile electrons down to
atomic length scales. The situation is, however, far more interesting on
the insulating side of disorder-driven metal-insulator transitions
\cite{dobrosavljevic:12}, where screening is suppressed due to charge
localization. Here, the unscreened Coulomb interaction leads to the
opening of the ``Coulomb gap'' in the electronic density of states, as
first pointed out in pioneering works of Pollack \cite{pollak:70}, as
well as Efros and Shklovskii (ES). The ES theory
\cite{efros:75,shklovskii:88} predicts a universal form of the Coulomb
gap, and explains how its existence modifies hopping transport
\cite{shklovskii:88} in disordered insulators, consistent with numerous
experiments \cite{lee:85}. Early work also revealed that Coulomb
interactions in disordered insulators generally contribute to the
formation of an extensive number of metastable states, i.e., the
formation of the Coulomb glass (CG) \cite{davies:82,davies:84,lee:88}.
In subsequent work, various aspects of glassy behavior of the CG were
explored theoretically
\cite{grannan:93,pastor:99,pastor:02,dobrosavljevic:03,mueller:04,
grempel:04,pankov:05,kolton:05,mueller:07,surer:09} and
experimentally
\cite{benchorin:93,ovadyahu:97,vaknin:00-ea,bogdanovich:02,
vaknin:02,orlyanchik:04,popovic:04,ovadyahu:06,jaroszynski:06,jaroszynski:07,raicevic:08,raicevic:11,lin:12}.

More  recent progress followed with the formulation of analytical
theories of the
CG \cite{pastor:99,pastor:02,dobrosavljevic:03,mueller:04,pankov:05,mueller:07}
which adapted Parisi's replica
methods \cite{parisi:83,rammal:86,mezard:87,young:98} for spin glasses to
disordered Coulomb systems. These theories find a Coulomb gap of the
same universal form as predicted by the ES theory, but this behavior
emerges only within the low-temperature glassy phase (displaying replica
symmetry breaking). Within this mean-field picture, the universality of
the Coulomb gap, as well as the saturation of the appropriate stability
bound, can be directly traced back to the ``marginal stability'' of the
entire glassy phase \cite{pastor:99}. In physical terms, the marginal
stability reflects the emergence of ``replicons,'' soft (gapless)
collective excitations involving simultaneous rearrangements of many
electrons. If such soft excitations indeed characterize the Coulomb
glass, they should also govern the physical response to various weak
perturbations (e.g.m the external electric fields), perhaps leading to
large-scale avalanches.  Precisely such behavior has already been
established \cite{pazmandi:99,andresen:13} for infinite-range spin-glass
models, leading to scale-free avalanches characterizing an entire
manifold of metastable states.  Despite the successes of the mean-field
approach, its applicability to finite space dimensions remains the
subject of much controversy and
debate \cite{young:04,katzgraber:05c,boettcher:07a,joerg:08a,katzgraber:09b,larson:13}.
Furthermore, a computational search for a finite-temperature glass
transition in the CG in two and three space dimensions has remained
inconclusive \cite{grempel:04,kolton:05,surer:09}. To shed
additional light on the nature of excitations in the CG, and further
test the mean-field ideas, it is therefore useful to examine the
stability of the low-lying metastable states by external electric
fields.

In this work, we investigate the out-of-equilibrium behavior of a
three-dimensional Coulomb glass at zero temperature and study the
hopping and total charge displacement avalanches triggered by increasing an
externally-applied electric field.  Previous work on avalanches in the
CG in three space dimensions done by Palassini and
Goethe \cite{palassini:12}, which trigger avalanches via dipole
excitations or charge insertions, find scale-free behavior for
long-range hopping dynamics, but when hopping is bounded by a finite
fixed range they do not find any scale-free avalanches. Because physical
electrons rearrange themselves by finite-range hopping it is of interest
to search for a scale-free behavior in the CG for bounded hopping
dynamics by other means. 

Here we study the CG with
particle-number-conserving short-range hopping, by ``adiabatically''
increasing an external electric field up to a depinning electric field
$\mathcal{E}_{\text{dp}}$ that separates the steady current state from
just finite electron rearrangements as a reaction to the external field.
We find that scale-free avalanches arise in the Coulomb glass when the
electric field is close to $\mathcal{E}_{\text{dp}}$.  To
emphasize the role played by the long-range Coulomb interactions we
repeat our simulations for an equivalent short-range interacting model.
In this case we still find a sharply defined depinning transition, but 
a completely different form for the critical behavior. Here we 
do not find any scale-free avalanches, in dramatic contrast
to the behavior of the CG model. 

The outline of this paper is as follows. Section \ref{sec:model}
describes the model, followed by a description of the used numerical
procedure in Sec.~\ref{sec:coulomb:algorithm}.  Measured quantities are
introduced in Sec.~\ref{sec:coulomb:observables}, followed by results
presented in Sec.~\ref{sec:results}.

\section{Model}
\label{sec:model}

The Coulomb glass Hamiltonian (in dimensionless units) is given
by \cite{efros:75}
\begin{align}
\mathcal{H} & =\frac{1}{2}\sum_{i\neq j}\left(n_{i}-K\right)
\frac{1}{\left|\textbf{ r}_{i}-\textbf{r}_{j}\right|}\left(n_{j}-K\right)
+\sum_{i}n_{i}\varphi_{i}\;,
\label{eq:coulomb:CGHamiltonian}
\end{align}
where $n_{i}$ is the electron number at site $i$, $K$ is the filling
factor, $\mathbf{r}_{i}$ is the coordinate of site $i$, and $\varphi_{i}$
a randomly-distributed on-site energy. For a charge neutral system,
i.e., $K=1/2$, in a constant external electric field $\mathcal{E}$ in
$x$-direction, Eq.~\eqref{eq:coulomb:CGHamiltonian} can be rewritten in
an Ising spin formulation by setting \cite{davies:82} $S_{i}=2n_{i}-1$
($S_i\in \{\pm 1\}$ an Ising spin variable)
\begin{align}
\mathcal{H} & =\frac{1}{4}\sum_{i<j}J_{ij}S_{i}S_{j}+\sum_{i}S_{i}
\left(\Phi_{i}+V_{i}\right)\,,
\label{eq:IsingLikeHamiltonian}
\end{align}
where the electric potential is $V_{i}=-\mathcal{E}x_{i}$ and $x_{i}$
is the $x$-position of spin $i$. This form of the Hamiltonian with
$\mathcal{E}=0$ is of a random-field Ising model with long-range
antiferromagnetic interactions given by 
\begin{align}
J_{ij}=\frac{1}{\left|\textbf{r}_{i}-\textbf{r}_{j}\right|}.
\label{eq:coulomb:interactions}
\end{align}
The site energy $\Phi_{i}=\varphi_{i}/2$ is sampled from a Gaussian
distribution with zero mean and standard deviation $\sigma=0.5$.  To
keep the dynamics of the two models identical it is necessary to
constrain the Ising-like Hamiltonian in
Eq.~\eqref{eq:IsingLikeHamiltonian} to have a constant magnetization
($m=0$ for $K=1/2$) at all times.  This is accomplished by using
magnetization-conserving Kawasaki dynamics \cite{newman:99}.

The corresponding short-range model (SR) is given by the same
Hamiltonian in Eq.~\eqref{eq:IsingLikeHamiltonian}, but with long-range
interactions replaced by nearest-neighbor interactions (on a cubic
lattice) of the form
\begin{align}
J_{ij} = \left\{ 
\begin{array}{ll}
1 & \quad
\text{if \ensuremath{i\;}and\;\ \ensuremath{j}\;\ are\;\ nearest\;\ neighbors,}\\
0 & \quad\text{otherwise}.
\end{array}\right.
\label{eq:coulomb:sh_interactions}
\end{align}

\subsection{Determination of the initial configurations}
\label{sec:coulomb:DOS} 

In our simulations, we need to generate stable initial configurations of
the system. In this context, ``stable'' refers to stable towards single
nearest-neighbor electron hopping.  We implement this procedure for both
the CG and the SR model. In order to have an initial configuration with
a Coulomb gap and track its dependence on the electric field, we compute
pseudo-ground-state configurations using jaded extremal optimization
(JEO) \cite{middleton:04}.

The single-particle density of states (DOS) of a classical Coulomb
system is given by
\begin{align}
\rho\left(E\right)=\left\langle 
\frac{1}{N}\sum_{i}\delta\left(E-E_{i}\right)
\right\rangle ,
\end{align}
where the local single-particle energy is given by 
\begin{align}
E_{i} & =\frac{1}{2}\sum_{j}J_{ij}S_{i}+2\Phi_{i}i
=\sum_{j}\left(n_{i} - \frac{1}{2}\right)J_{ij}+\varphi_{i},
\label{eq:coulomb:single_particle_energy}
\end{align}
and the average $\langle \cdots \rangle$ is performed both over thermal
fluctuations and disorder instances. The ground state of the CG is well
known to display a Coulomb gap \cite{efros:75} in the DOS at the Fermi
energy, which gradually fills up when temperature is
increased \cite{davies:82,davies:84,grannan:93,sarvestani:95,surer:09}.

\begin{figure}
\includegraphics[height=0.23\textheight]{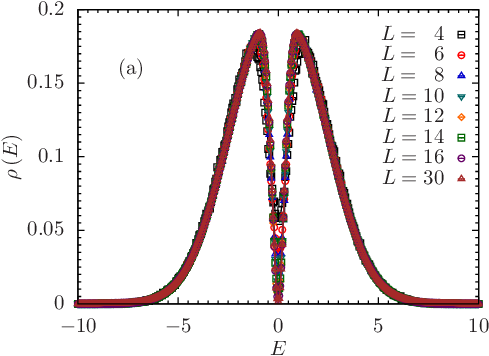} 
\includegraphics[height=0.23\textheight]{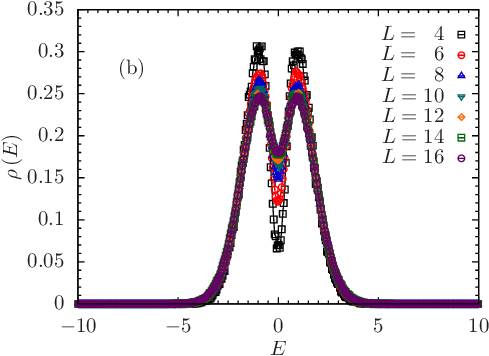}
\vspace*{-0.2cm}
\caption{(Color online)
Density of states for the three-dimensional CG of (a) the starting
pseudo-ground-state configurations and (b) over a range of different
electric potentials $0.5<\mathcal{E}<0.6$. Both distributions show a
clear dip for $E = 0$, suggesting that the states computed using JEO are
indeed close to the true ground state of the system. Data averaged over
$2500$ -- $10000$ disorder instances, depending on the size size of the
system (see Table \ref{tab:simparams}).
\label{fig:coulomb:DOS_CG} 
}
\end{figure}

For the CG, we can empirically check how ``far'' or ``close'' a given
configuration is from the ground state by examining the form of the
DOS. Depending on the depth of the Coulomb gap, we can argue whether the
configurations are close or far from their respective ground state. The
SR ground states do not have a Coulomb gap \cite{boettcher:07a}, but have
a ``dip'' at the Fermi energy that converges to a finite value in the
thermodynamic limit. Again, we can empirically check if we have a good
approximation of the ground state by studying at the DOS distribution.
In Fig.~\ref{fig:coulomb:DOS_CG}(a), we show the DOS of the CG using the
pseudo ground states for all simulated linear system sizes $L$ (the
systems have $N = L^3$ spins). The occupation at $E=0$ is very close to
zero, showing that the configurations found using JEO are not far from
the true ground state. In Fig.~\ref{fig:coulomb:DOS_CG}(b), we show the
DOS of the CG at electric fields $0.5<\mathcal{E}<0.6$.  The data
suggest that we are further away from a ground-state configuration,
however, a pronounced gap in the DOS is still visible.  The
configurations for the SR model found by the JEO algorithm are likewise
not far from the ground state (not shown).

\section{Numerical Details}
\label{sec:nums}

\subsection{Algorithm}
\label{sec:coulomb:algorithm} 

\begin{figure}[b]
\includegraphics[width=0.15\textwidth]{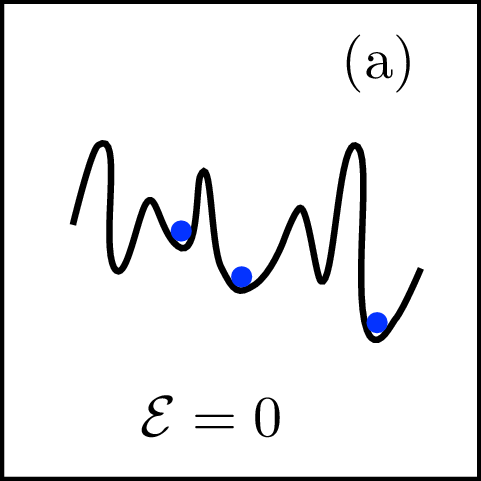}
\includegraphics[width=0.15\textwidth]{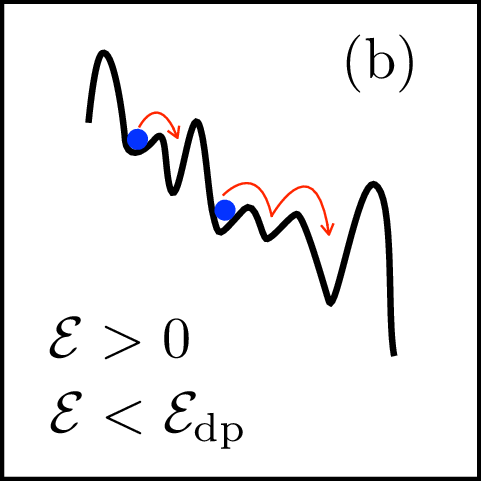}
\includegraphics[width=0.15\textwidth]{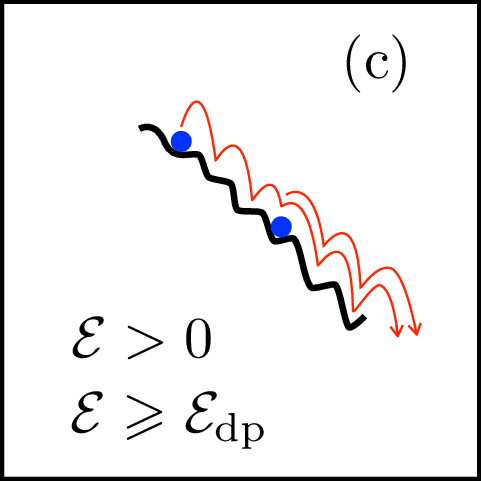}
\caption{(Color online)
Sketch of the site-dependent random potential landscape felt by the
electrons (blue circles) at different electric field strengths: (a)
$\mathcal{E}=0$, (b) $0<\mathcal{E}<\mathcal{E}_{\text{dp}}$, and (c)
$\mathcal{E}_{\text{dp}}\leqslant\mathcal{E}$. (a) Stable configuration
of electrons at $\mathcal{E}=0$. (b) The electric field effectively
tilts the potential. At electric fields
$0 < \mathcal{E}<\mathcal{E}_{\text{dp}}$ the electrons just rearrange as a
reaction to the field. (c) The electric field $\mathcal{E} \geqslant
\mathcal{E}_{\text{dp}}$ further tilts the potential to a point where a
steady current is induced.
\label{fig:coulomb:sketch} 
}
\end{figure}

For the description of the algorithm, we introduce a stability criterion,
which for an electron ($S_{i}=1$) or a vacancy ($S_{i}=-1$) at a given
site is given by
\begin{align}
\left(E_{i}+V_{i}\right)\cdot S_{i}<0\;\; & 
\rightarrow\;\;\text{stable,}\label{eq:stability}\\
\left(E_{i}+V_{i}\right)\cdot S_{i}>0\;\; & 
\rightarrow\;\;\text{unstable.}
\label{eq:unstability}
\end{align}
For each pseudo-ground state generated via JEO [see
Fig.~\ref{fig:coulomb:DOS_CG}(a)], we proceed as follows.
\begin{enumerate}

\item Select the least stable electron with one nearest-neighbor hole in
the opposite direction of the electric field. 
\label{eq:coulomb:step1}

\item Apply an electric field $\mathcal{E}$ just strong enough to
destabilize the selected electron, such that it will hop to the
neighboring hole.

\item Recompute all single-particle energies given by
Eq.~\eqref{eq:coulomb:single_particle_energy}, and select the most
unstable electron that minimizes the total energy by hopping to one of
its neighboring holes. If there are no unstable electrons or an energy
minimization is not possible, go to step~\ref{eq:coulomb:step1}.
\label{eq:coulomb:step3}

\item Perform the electron-hole hopping that minimizes the energy; go to
step~\ref{eq:coulomb:step3}.  
\label{eq:coulomb:step4}

\end{enumerate}
The careful reader will have noticed that the above procedure is in fact
an infinite loop stuck between steps \ref{eq:coulomb:step3} and
\ref{eq:coulomb:step4} when a certain electric field threshold
$\mathcal{E}\geqslant\mathcal{E}_{\text{dp}}$ is reached. This
electric field threshold is the depinning field of the system, which
separates two regions: Below $\mathcal{E}_{\text{dp}}$ there are only
short charge displacement pulses due to the rearrangement of the electrons as a
response to the external electric field, and above it there is a steady
current. A sketch of the different scenarios is shown in
Fig.~\ref{fig:coulomb:sketch}. The infinite loop between step
\ref{eq:coulomb:step3} and step \ref{eq:coulomb:step4} is the steady
current flowing through the system. Since we are interested in the
number of times step \ref{eq:coulomb:step3} and step
\ref{eq:coulomb:step4} are repeated at each $\mathcal{E}$-field (this,
in turn, yields the avalanche size $n$) before we reach the depinning
field, we artificially stop the process if the avalanche size surpasses
a given number $n_{\text{steady}}=2N$, where $N$ is the total number of
sites of the system. Note that $n_{\text{steady}}$ is much larger than
the maximal avalanche size measured for
$\mathcal{E}<\mathcal{E}_{\text{dp}}$ for a given system size $L$.

To cope with the long-range Coulomb interactions between the electrons
we use the Ewald summation method \cite{wang:01b}. Furthermore, the
applied electric field is periodic to avoid an electron pileup at the
edge of the system.  The simulation parameters are listed in
Tab.~\ref{tab:simparams}.

\begin{table}
\caption{
Parameters of the simulation: For the Coulomb glass (CG) and the
short-range model (SR) we study systems of $N = L^3$ spins close to the
ground state and compute the different distributions over $N_{\rm sa}$
disorder samples for different applied electric fields $\mathcal{E}$.
\label{tab:simparams}
}
{\scriptsize
\begin{tabular*}{\columnwidth}{@{\extracolsep{\fill}} c r r}
\hline
\hline
$\text{model}$ & $L$ & $N_{\rm sa}$ \\
\hline
CG	     & $4$ &    $8\,000$ \\
CG	     & $6$ &    $9\,000$ \\
CG	     & $8$ &    $6\,500$ \\
CG	     & $10$ &    $5\,000$ \\
CG	     & $12$ &    $4\,000$ \\
CG	     & $14$ &    $9\,000$ \\
CG	     & $16$ &    $4\,000$ \\
CG	     & $30$ &    $2\,500$ \\
\\[-0.75em]
\hline
\\[-0.75em]
SR	     & $4$ &    $12\,000$ \\
SR	     & $8$ &    $14\,400$ \\
SR	     & $16$ &   $10\,200$ \\
SR	     & $24$ &    $9\,500$ \\
SR	     & $32$ &    $7\,400$ \\
SR	     & $48$ &    $2\,500$ \\
SR	     & $64$ &    $\;\,700$ \\
\hline
\hline
\end{tabular*}
}
\end{table}

\subsection{Measured observables and statistical data analysis}
\label{sec:coulomb:observables} 

\begin{figure}
\includegraphics[height=0.23\textheight]{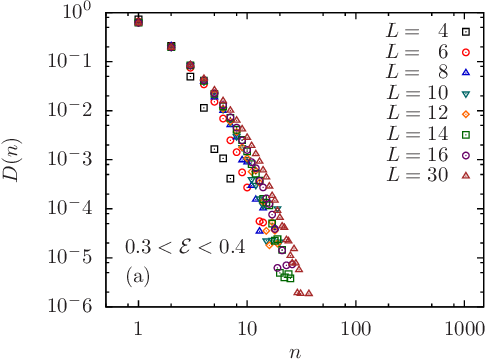}
\includegraphics[height=0.23\textheight]{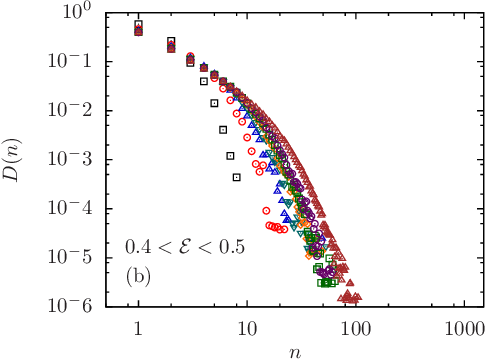}
\includegraphics[height=0.23\textheight]{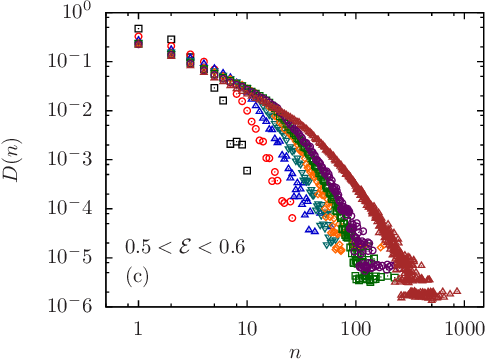}
\includegraphics[height=0.23\textheight]{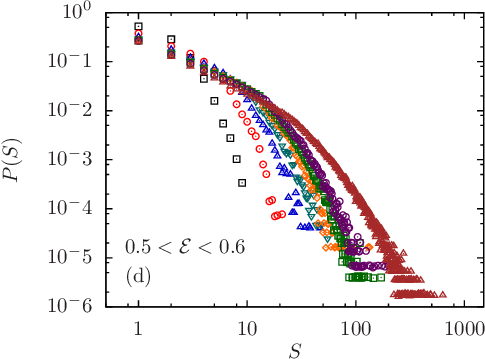}
\vspace*{-0.2cm}
 \caption{(Color online)
(a), (b) and (c) show electron-hole avalanche distributions
$D(n)$ of the CG at electric field ranges between $0.3<\mathcal{E}<0.6$.
Scale-free avalanches emerge as $\mathcal{E}$ approaches
$\mathcal{E}_{\text{dp}} \approx 0.603(5)$. (a) $0.3<\mathcal{E}<0.4$,
(b) $0.4<\mathcal{E}<0.5$ and (c) $0.5<\mathcal{E}<0.6$. Note that only
close to the depinning electric field
$\mathcal{E}_{\text{dp}}\approx0.603(5)$ scale-free avalanches, i.e.,
power-law distributions of avalanche sizes, emerge.  (d) Distribution of
charge displacement spikes (avalanches) $P(S)$ of the CG for
$0.5<\mathcal{E}<0.6$.
\label{fig:coulomb:avalanches_CG}
}
\end{figure}

\begin{figure}[t]
\includegraphics[height=0.23\textheight]{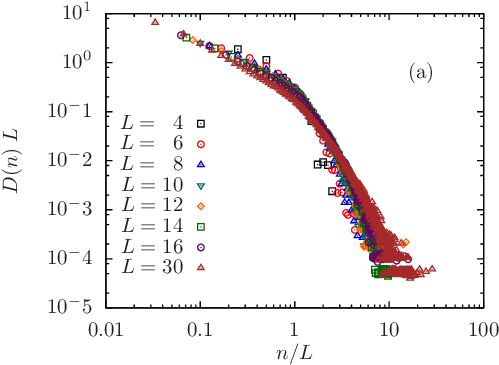}
\includegraphics[height=0.23\textheight]{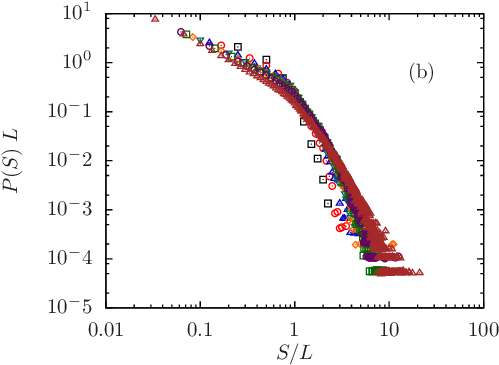}
\vspace*{-0.2cm}
\caption{(Color online) 
Finite-size scaling data collapse of the electron avalanche
distributions $D(n)$ (a) according to
Eq.~\eqref{eq:scaling_collapse_1} with $0.5<\mathcal{E}<0.6$, i.e.,
close to  $\mathcal{E}_{\text{dp}}$.  For the largest system sizes the
data seem to collapse well. (b) shows a data collapse of the
charge displacement distributions $P(S)$ according to
Eq.~\eqref{eq:scaling_collapse_2} with $0.5<\mathcal{E}<0.6$.  Again,
the data scale well. Note that the symbols used are the same as in 
(a).
\label{fig:coulomb:scaling_CG} 
}
\end{figure}

\begin{figure}
\includegraphics[height=0.23\textheight]{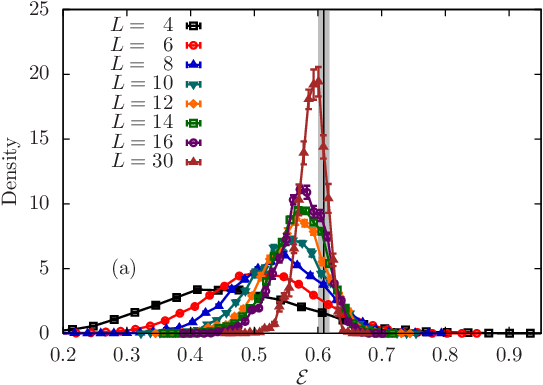} 
\includegraphics[height=0.23\textheight]{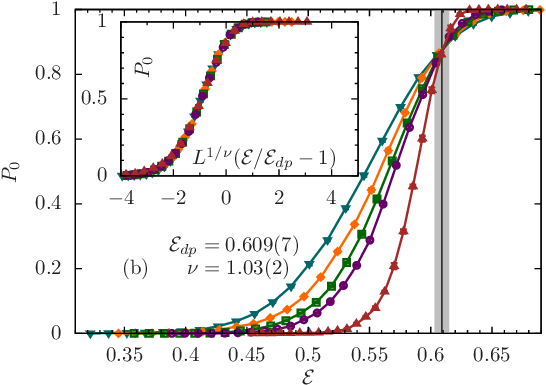} \vspace*{-0.2cm}
\caption{(Color online) 
(a) Depinning distributions for all system sizes studied of the CG model. The
vertical line represents the estimated depinning field $\mathcal{E}_{dp}$.  (b)
The cumulative distribution function of the depinning field for linear system
sizes $L\geq10$. The curves for different system sizes cross a the depinning
field value.  The inset is a data collapse assuming the universal function
$P_0$ scales as Eq.~(\ref{eq:scale}).
\label{fig:coulomb:depinning_CG} }
\end{figure}

\begin{figure}
\includegraphics[height=0.23\textheight]{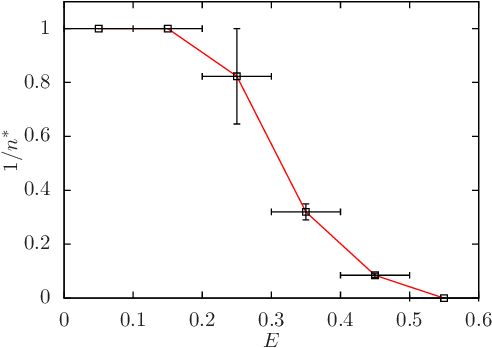}
\vspace*{-0.2cm}
\caption{(Color online) 
Characteristic avalanche size $n^*$, computed using Eq.~(\ref{eq:ansatz}),
as a function of the applied field $\mathcal{E}$ for the CG model.  As the
field increases, the inverse of the characteristic avalanche size $1/n^*$
decreases until at the depinning field $\mathcal{E}_{\text{dp}}$ it becomes
zero, i.e., $n^*\left(\mathcal{E}_{\text{dp}}\right)\rightarrow \infty$ (the
line is a guide to the eye).
\label{fig:coulomb:e_vs_n_star} 
}
\end{figure}

\begin{figure}
\includegraphics[height=0.23\textheight]{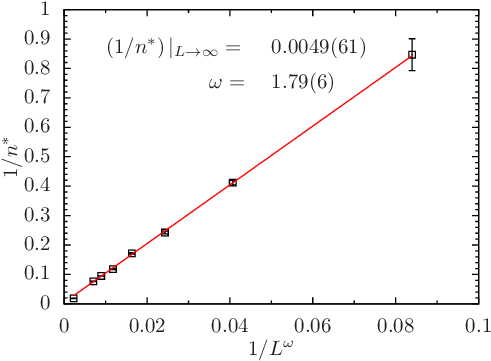}
\vspace*{-0.2cm}
\caption{(Color online) 
Thermodynamic limit extrapolation of the characteristic avalanche size
$n^{*}$ for the CG model in an electric field $0.5<\mathcal{E}<0.6$
close the depinning field $\mathcal{E}_{\text{dp}}=0.603(5)$ .  We fit
the data to Eq.~\eqref{eq:ansatz} with $1/n_{\infty}^{*}$, $a$, and
$\omega$ parameters. An optimal fit gives $1/n^{*}_{\text{CG}} =
0.0049(61)$ [$\omega = 1.79(6)$] with a quality-of-fit
probability \cite{press:95} $Q = 0.994$. Note that fixing
$1/n_{\infty}^{*} = 0$ gives $Q = 0.998$.  This means that
$n_{\infty}^{*} = \infty$, i.e., the presence of scale-free avalanches
in this electric field regime.
\label{fig:coulomb:extrapolation_CG} 
}
\end{figure}

\begin{figure}
\includegraphics[height=0.23\textheight]{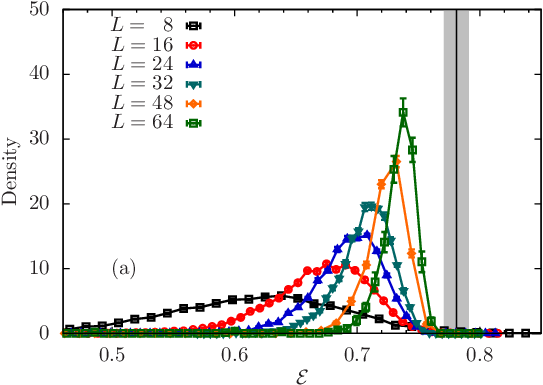} 
\includegraphics[height=0.23\textheight]{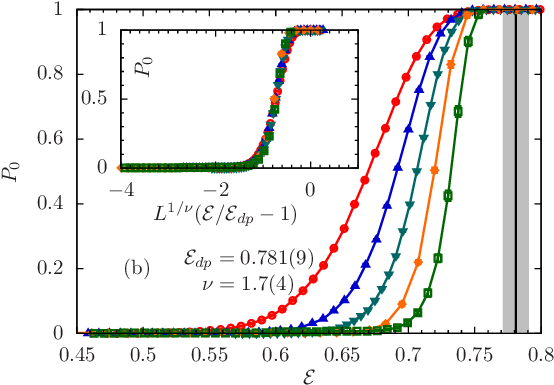} \vspace*{-0.2cm}
\caption{(Color online) 
(a) Depinning distributions for all system sizes studied of the SR model. The
vertical line represents the estimated depinning field $\mathcal{E}_{dp}$.  (b)
The cumulative distribution function of the depinning field for linear system
sizes $L\geq16$. The curves for different system sizes cross a the depinning
field value. Here the crossing seems to happen in a region where it is not
possible to distinguish it. Nevertheless, as seen in the inset, assuming the
universal function $P_0$ scales as shown in Eq.~(\ref{eq:scale}) the data
collapse is satisfactory.
\label{fig:coulomb:depinning_SR} }
\end{figure}

At each increase of $\mathcal{E}$ we count the number of electrons $n$ that
hopped and the total charge displacement $S$ in the direction of the applied
electric field. Using these data, we compute their distributions $D(n)$ and
$P(S)$, respectively (see, for example, Fig.~\ref{fig:coulomb:avalanches_CG}).
To determine the depinning field $\mathcal{E}_{\text{dp}}$ we compute the
cumulative distribution function $P_0(L,\mathcal{E})$ of the depinning
distributions which gives the probability whether a randomly picked sample is
in the pinned or depinned state for a given system size and at a given field.
We perform a finite-size scaling assuming that the function $P_0$ has a
universal form \cite{dong:93,newman:00,xi:15}
\begin{equation}
P_0\sim\tilde\Phi[ L^{1/\nu}(\mathcal{E}/\mathcal{E}_{dp} - 1 )]
\label{eq:scale}
\end{equation}
[see Fig.~\ref{fig:coulomb:depinning_CG}(b) and
Fig.~\ref{fig:coulomb:depinning_SR}(b) for the CG model and the SR
model, respectively], which gives us an estimate of the depinning field.
Note that the depinning field is defined as the typical electric field
necessary to induce a continuous current for a given system size, i.e,
for $\mathcal{E}<\mathcal{E}_{\text{dp}}$, the system just rearranges its
electron configuration by electron hopping, whereas for
$\mathcal{E}>\mathcal{E}_{\text{dp}}$, the field induces a steady
current.

In addition, we define the characteristic avalanche size $n^{*}$ of the
system by fitting the exponential tail of the avalanche distributions
$D(n)$ to an exponential function $\sim\exp(-n/n^{*})$.  For each system
size $L$, we thus obtain a characteristic avalanche size $n^{*}(L)$.  To
estimate the value of $n_{\infty}^{*}$ in the thermodynamic limit, we do
an extrapolation of $n_{L\rightarrow\infty}^{*}$ by using the following
functional ansatz:
\begin{equation}
1/n_{L}^{*}=1/n_{\infty}^{*} + a/L^{\omega}\;, 
\label{eq:ansatz}
\end{equation}
where $\omega$, $a$, and $n_{\infty}^{*}$ are fitting parameters.

Finally, we also monitor the DOS as a function of the applied electric
field $\mathcal{E}$. For example, Fig.~\ref{fig:coulomb:DOS_CG}(b)
shows the density of states at an electric field range of
$0.5<\mathcal{E}<0.6$.

Different finite-size scaling Ans\"{a}tze have been
attempted \cite{pazmandi:99} to scale the $D(n)$ and $P(S)$ data without
yielding any satisfactory results.  We therefore empirically re-sized
the avalanche curves without making any a priori assumptions.
Interestingly, the following scaling ansatz showed good
results:
\begin{align}
D & =\frac{1}{L}\; d\left(n/L\right)\label{eq:scaling_collapse_1}\\
P & =\frac{1}{L}\; p\left(S/L\right)\;,\label{eq:scaling_collapse_2}
\end{align}
where $d\left(n/L\right)$ and $p\left(S/L\right)$ in
Eqs.~\eqref{eq:scaling_collapse_1} and
\eqref{eq:scaling_collapse_2}, respectively, are universal functions.

\begin{figure}
\includegraphics[height=0.23\textheight]{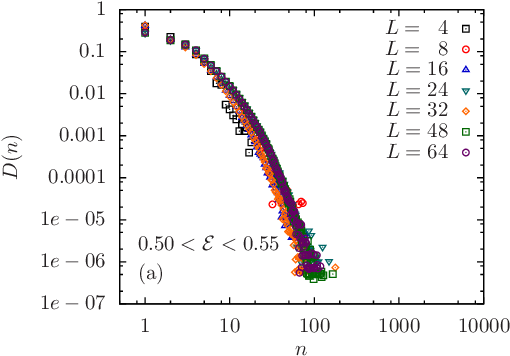}
\includegraphics[height=0.23\textheight]{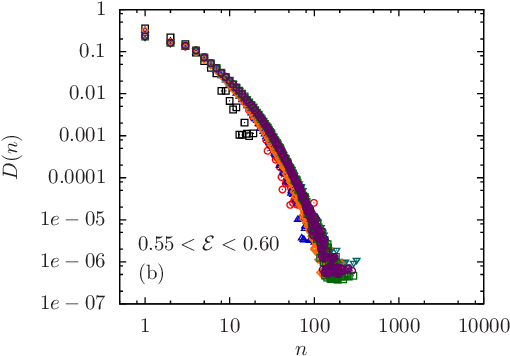}
\includegraphics[height=0.23\textheight]{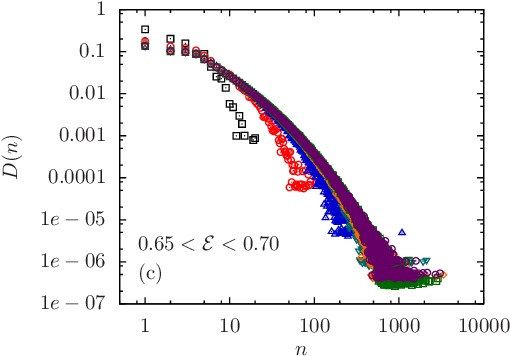}
\includegraphics[height=0.23\textheight]{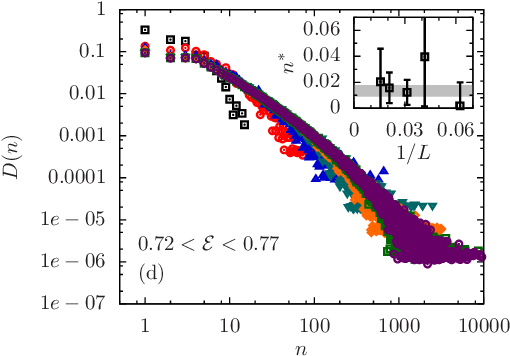}
\vspace*{-0.2cm}
\caption{(Color online) 
Spin avalanches $D\left(n\right)$ of the SR model at different electric field
ranges: (a) $0.50<\mathcal{E}<0.55$, (b) $0.55<\mathcal{E}<0.60$, (c)
$0.65<\mathcal{E}<0.70$ and (d) $0.72<\mathcal{E}<0.77$. Even for
$\mathcal{E}\approx\mathcal{E}_{\text{dp}}^{\text{SR}}$ (d) there is no
sign of scale-free avalanches. The inset in (d) shows the $n^*$ estimates for
system sizes $L\geqslant 16$ and the gray horizontal line is their mean value.
\label{fig:coulomb:avalanches_SR} 
}
\end{figure}

\section{Results}
\label{sec:results}

Figure \ref{fig:coulomb:avalanches_CG} shows electron hop, as well as
total charge displacement avalanche distributions for the CG for
different ranges of the electric field $\mathcal{E}$. The field
$\mathcal{E}$ is increased in the different panels from top to bottom.
Figures \ref{fig:coulomb:avalanches_CG}(a) --
\ref{fig:coulomb:avalanches_CG}(c) show how the avalanche sizes
progressively become system spanning, i.e., when $\mathcal{E} \approx
\mathcal{E}_{\text{dp}}$ [as is the case in
Fig.~\ref{fig:coulomb:avalanches_CG}(c)] avalanche size distributions
become power laws.  As the field reaches $\mathcal{E}_{\text{dp}}$ a
hunch in the curves emerges separating a power-law region from an
exponential cutoff, for the measured avalanches distribution
$D\left(n\right)$.  Figure~\ref{fig:coulomb:e_vs_n_star} shows the
dependence of the inverse of the characteristic avalanche size $1/n^*$
as a function of the electric field $\mathcal{E}_{\text{dp}}$. We can
extract from the figure that the depinning field
$\mathcal{E}_{\text{dp}}$ lies somewhere around $\mathcal{E}\approx0.6$.
A precise estimate of the depinning field can be obtained by analyzing
the cumulative distribution function $P_0$ as shown in
Fig.~\ref{fig:coulomb:depinning_CG}(b). For the CG model we obtain
$\mathcal{E}{_\text{dp}} = 0.609(7)$.  In
Fig.~\ref{fig:coulomb:extrapolation_CG} we show an example of the
estimation of $n^*$ using Eq.~(\ref{eq:ansatz}) for a given field window
$0.5<\mathcal{E}<0.6$. Similar qualitative results are obtained for the
charge displacement distribution $P\left(S\right)$, as shown in
Fig.~\ref{fig:coulomb:avalanches_CG}(d). We attempt to scale the data
for the distributions $D(n)$ and $P(S)$ in
Fig.~\ref{fig:coulomb:scaling_CG}. The data scale well with no
adjustable parameters (especially for the larger system sizes) according
to Eqs.~\eqref{eq:scaling_collapse_1} and \eqref{eq:scaling_collapse_2}.

\begin{figure}
\includegraphics[height=0.23\textheight]{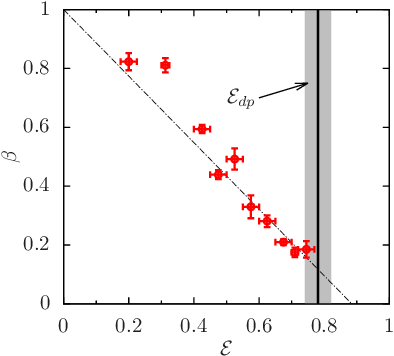} 
\caption{(Color online) Stretched exponential exponent $\beta$ of the SR model
as a function of the applied field $\mathcal{E}$. The exponent $\beta$
decreases monotonically with the field. The vertical line shows the estimated
depinning field $\mathcal{E}_{dp}$ and the dashed line is a guide to the eye.
\label{fig:SR:Beta_vs_E} }
\end{figure}

In addition, we study the total charge displacement distribution and
electron hop distribution as a function of the applied field for the SR
model, where the estimated depinning field is
$\mathcal{E}_{\text{dp}}=0.781(9)$ as seen in
Fig.~\ref{fig:coulomb:depinning_SR}. Electron avalanche distributions
are shown in Fig.~\ref{fig:coulomb:avalanches_SR}. For low fields, i.e.
$\mathcal{E}<0.5$, the characteristic avalanche size $n^*\left(
L\right)$ can be estimated analogously as for the CG model, i.e.,
fitting the tail to an exponential function and using
Eq.~(\ref{eq:ansatz}) to extrapolate to the thermodynamic limit. As for
the CG model at low fields, no system-spanning avalanches were found,
moreover no emergent avalanche size dependence is observed
[Fig.~\ref{fig:coulomb:avalanches_SR}(a)-(b)]. For fields closer to the
depinning field, i.e. $\mathcal{E} \gtrsim 0.5$, the exponential fitting
function [Eq.~(\ref{eq:ansatz})] gives unsatisfactory fitting results,
therefore we additionally fitted the distribution to a stretched
exponential function 
\begin{equation}
f(x)= a_L\;\exp[-\left(x/n^*_L\right)^{\beta_L}] . 
\label{eq:fx}
\end{equation}
The characteristic avalanche size
$n^*$ defined through the stretched exponential function is bounded in
the thermodynamic limit for all fields, especially close to the depinning
field: the inset of Fig.~\ref{fig:coulomb:avalanches_SR}(d) shows the
values of $n^*$ for the field window $0.72<\mathcal{E}<0.77$. The
stretched exponential exponent $\beta$ has a strong field dependence as
seen in Fig.~\ref{fig:SR:Beta_vs_E}. At low fields $\beta\approx 0.8$
and as the field increases it monotonically decreases to
$\beta\approx0.2$ \cite{comment:stretched_exp}.

We observe that the CG model and the SR model have a well defined
depinning field transition, but that they differ in the way they behave
close to $\mathcal{E}_{dp}$.  The CG model total charge displacement and
electron hop avalanche distributions close to the depinning field have a
power-law shape (with power-law exponent $\tau\approx -1$) with a
system-size dependent exponential cutoff. This finite-size effect
vanishes in the thermodynamic limit, revealing its scale-free behavior
at $\mathcal{E}_{dp}$. In clear contrast the SR model total charge
displacement and electron hop avalanche distributions show no signs of
scale-free avalanche behavior (power-law shape) close to
$\mathcal{E}_{dp}$ and are best described by a stretched exponential
function, which is defined by the exponent $\beta$ and the parameter
$n^*$. The exponent $\beta$ shows a strong field dependence; it
decreases monotonically as the field is increased, while $n^*$ does not
show any systematic system-size dependence at any field, not even close
to the depinning field. The different avalanche distributions in the SR
and CG models hint towards a different mechanism behind the depinning
transition.

\section{Conclusions} 
\label{sec:concl}

Our large-scale computational study of the Coulomb glass has
demonstrated that, under external electric fields and nearest-neighbor
particle-conserving hopping dynamics, scale-free avalanches only occur
in the vicinity of a characteristic depinning field
$\mathcal{E}_{\text{dp}}$. For small external electric fields, no large
avalanches are present, in agreement with the results of Palassini and
Goethe \cite{palassini:12}. For a short-range variation of the Coulomb
glass model we do not find any sign of scale-free avalanches, not even
close to the depinning electric field.  Furthermore, we find that the
initial Coulomb gap vanishes as the field is ramped up, suggesting that
it is not a generic feature on the hysteresis loop formed in an external
electric field. We empirically find a simple scaling ansatz to collapse
the avalanche and charge displacement distributions, reinforcing the
notion that the scale-free behavior of the CG emerges close to the
depinning electric field.

The scale-free behavior found in the CG is not a self-organized critical
(SOC) state, because an external parameter has to be
tuned \cite{bak:87,drossel:92,schenk:02,andresen:13}, namely the electric
field $\mathcal{E}$. Nevertheless, it is interesting to note the
difference between the CG and the SR model: in the former the
combination of the diverging number of neighbors and disorder results in
power-law distributions, which is not the case in the latter. This
behavior is very similar to that found for the three-dimensional
random-field Ising
model \cite{sethna:93,perkovic:95,perkovic:99,kuntz:98,sethna:04}, where
scale-free avalanches have been observed at a critical field strength
$h_{c}$.  These unexpected results for the Coulomb glass show that a
diverging number of neighbors is necessary but {\em not} sufficient in a
model Hamiltonian to show SOC behavior, and that the dynamics of a model
might play an important role for showing SOC (i.e., the order-parameter
conserving Kawasaki dynamics used here vs single-spin flip dynamics used
for the random-field Ising model).

Our results bring into question the validity of the mean-field picture
of the Coulomb
glass \cite{pastor:99,pastor:02,dobrosavljevic:03,mueller:04,pankov:05,mueller:07},
predicting extreme fragility of the ground state to external
perturbations.  However, the generic absence of SOC for avalanches
driven by a uniform electric field may be related to the fact that such
large avalanches {\em locally} violate charge neutrality. Other
dynamical perturbations may couple differently to the elementary
excitations and may perhaps serve as a more sensitive probe to the
proposed SOC nature of the CG ground state. This could be achieved by
applying external fields that do not directly couple to the {\em
uniform} charge density, such as varying the amplitude of the disorder
potential. Such or similar studies represent an opportunity to further
elucidate the long-standing mystery of the Coulomb glass, however,
exploring this exciting research direction remains a challenge for
future work.

\begin{acknowledgments}

We would like to thank R.~S.~Andrist for many discussions, as well as
Mauricio Andresen for providing the necessary motivation to complete
this project. V.D.~and Y.P.~were supported by the NSF (Grant
No.~DMR-1005751). G.T.Z.~was supported by the NSF (Grant
No.~DMR-1035468).  H.G.K.~acknowledges support from the NSF (Grant
No.~DMR-1151387) and would like to thank ETH Zurich for CPU time on the
Brutus cluster, as well as Aspall (Suffolk) for inspiration.  Part of
H.G.K.'s research is based upon work supported in part by the Office
of the Director of National Intelligence (ODNI), Intelligence Advanced
Research Projects Activity (IARPA), via MIT Lincoln Laboratory Air Force
Contract No.~FA8721-05-C-0002.  The views and conclusions contained
herein are those of the authors and should not be interpreted as
necessarily representing the official policies or endorsements, either
expressed or implied, of ODNI, IARPA, or the U.S.~Government.  The
U.S.~Government is authorized to reproduce and distribute reprints for
Governmental purpose notwithstanding any copyright annotation thereon.

\end{acknowledgments}

\bibliography{refs,comments}

\begin{thebibliography}{61}
\bibitem[{com()}]{comment:stretched_exp}
\bibinfo{note}{We also applied a stretched exponential fitting function to the
  CG model avalanche distributions. The stretched exponent $\beta$ for the CG
  model are in good agreement with an exponential function for fields
  $\mathcal{E}\lesssim 0.5$. Close to $\mathcal{E}_{dp}$ the exponent $\beta$
  is $0.5\lesssim \beta\lesssim 0.8$. Nevertheless, a pure exponential function
  still gives satisfactory fitting quality values.}
\end{thebibliography}

\end{document}